\documentclass[runningheads]{llncs}
\usepackage{geometry}
\usepackage{graphicx}
\usepackage{amsmath}
\usepackage{amssymb}
\usepackage{mhchem}
\usepackage{physics}
\usepackage{subcaption}
\usepackage{hyperref}
\begin{document}
\title{Digestive System Dynamics in Molecular Communication Perspectives
\thanks{Supported by {\it VistaMilk} Research Center, Ireland.}}
%
%
\author{Dixon Vimalajeewa\inst{1}\orcidID{0000-0001-6794-4776} \and
Sasitharan Balasubramaniam\inst{1}\orcidID{0000-0003-4022-3615} 
}
\authorrunning{D. Vimalajeewa et al.}
%
\institute{Walton Institute, Waterford Institute of Technology, Waterford, Ireland 
\email{ \{dixon.vimalajeewa, sasi.bala\}@waltoninstitute.ie}\\
\url{https://waltoninstitute.ie/} 
}
\maketitle              
\begin{abstract}
Consumption of food in excess of the required optimal nutritional requirements has already resulted in a global crisis and this is from the perspective of human health, such as obesity, as well as food waste and sustainability. In order to minimize the impact of these issues, there is a need to develop novel innovative and effective solutions that can optimally match the food consumption to the demand. This requires accurate understanding of the food digestion dynamics and its impact on each individual's physiological characteristics. This study proposes a model to characterize digestive system dynamics by using concepts from the field of Molecular Communications (MC), and this includes integrating advection-diffusion and reaction mechanisms and its role in characterizing the digestion process as a communication system. The model is then used to explore starch digestion dynamics by using  communication system metrics such as delay and path loss. Our simulations found that the long gastric emptying time increases the delay in starch digestion and in turn the glucose production and absorption into the blood stream. At the same time, the enzyme activity on the hydrolyzed starch  directly impacts the path loss, as higher reaction rates and lower half saturation concentration of starch results in lower path loss. Our work can lead to provide insights formulated for each individuals by creating a digital twin digestion model. 
\end{abstract}
\section{Introduction}
The United Nations (UN) aims to halve the per capita global food waste at the retail and consumer levels by 2030 as $\sim 1/3$ of food produced globally for human consumption is wasted \cite{1}. Metabolic food waste and food consumption in excess of the optimal nutritional requirements, uses valuable agricultural resources and results in critical health issues such as obesity. Besides, obesity in humans is also associated with $\sim 20\%$ greenhouse gas emissions that is relative to the normal-weight state estimated from increased food intake, aerobic metabolism and fossil fuel use for transportation \cite{2}. These facts emphasis the need for innovative solutions to overcome such challenges and achieve the goals set by the UN. 

Facilitating to set up more sustainable food consumption habits is one innovative solutions that is being widely investigated, where the optimal proportion of nutrition can be allocated to each individual based on their health condition as well as physiological setup and lifestyle. This facilitates consumers to design food that suits their personal nutritional needs, by considering factors such as age, gender, and physical fitness. Using this practice will  help improve global human health and reduce food waste in a number of ways. For instance, setting up proper dietary plan and exercise routines to maintain optimal body weight. From a digital modeling perspective, digestive system dynamics can reveal valuable information which can be used to decide suitable food based on suitability to individual personal traits and health benefits \cite{4}. Exploring digestive system dynamics can thus contribute to designing frameworks for setting up more sustainable food consumption habits, which will result in  a more population-oriented food production and supply chain in the near future.


With the focus on expanding knowledge about the digestive system dynamics, this study aims to looks at its functionality from a different angle compared to the existing approaches, by proposing a mathematical model that can be used to a Molecular Communication (MC) system representation. A majority of previous studies has considered the use of computational models to characterize digestive system dynamics by using its physio-chemical properties. For instance, the study \cite{20} provides different modeling methodologies, particularly compartmental models to characterize gastrointestinal tract dynamics during digestion. The study \cite{5} also presents a compartment modeling approach using chemical kinetics in the context of drug administration, while authors in  \cite{6} proposes a set of computational models considering different properties of food which include completely soluble, non-soluble and non-degradable types. In addition, physiologically based pharmokinetics models have also been used to explore the effects of different physiological parameters such as age and disease status and then use derived insights for drug discovery and development \cite{19}. This study takes a different direction by developing a computational model that uses concepts from MC, and used to mapped its functionality from a communication theory perspective. The recent introduction of MC has provided a new mechanism of characterizing biological cellular communication, and also facilitates new approaches of using engineered bionanoscience to control the communication process. From both pharmaceutics and food sciences perspectives, this can lead to development of innovative solutions for various critical issues such as targeted drug delivery for various nutritional diseases \cite{7} and controlled delivery of nutrients \cite{8}.

In the model building process, three main stages of the digestive process are modeled as a MC system  as illustrated in  Figure \ref{Digest}.  First, the stomach releases mechanically and chemically broken down food mixed with gastric juice (\emph{digesta}) into the small intestine (this study terms this as the \emph{small intestinal (SI) tract }). In the second stage, the digesta reacts with different enzymes available in the SI tract and converts it into an absorbable form (nutrients) while traveling along the SI tract. Finally, in the third stage, the SI wall selectively absorbs them into the circulatory system. These three stages are then mapped to an advection-diffusion, and reaction based MC system \cite{15} (further explanation will be provided in the following section). The model is derived by considering the digestion of starch process. Exploring starch digestion dynamics is very important as it is the main source of carbohydrates in our diet and plays a crucial role as a source of energy. Carbohydrates supports fat metabolism as well as effective use of dietary proteins \cite{9}. However, excess consumption of carbohydrates along with dietary fat can contribute towards critical health issues such as obesity.  Thus, broader knowledge about carbohydrate digestion dynamics can help in diabetes and obesity management \cite{16}. This study takes into account  the starch digestion and absorption of glucose into the blood as a advection-diffusion and reaction based MC system. This is followed by understanding the  impact of different parameters such as half-gastric emptying time (i.e., time to empty half of digesta volume) on the digestive system dynamics. Finally, performance metrics used in communication  theory such as delay and path loss, are used to evaluate the performances of the starch digestion system.  

The reminder of the study is organized as follows. Next section \ref{model} presents the system model and this is followed by the results derived through simulation studies using Python software simulation in section \ref{res}. The importance of the results and their implications along with further studies are discussed in section \ref{dis} and finally, section \ref{con} concludes the paper. 
%
%
\section{System Model}\label{model}
This section first gives a brief overview about the food digestion process and its mapping to a MC system. Next, a transport model for nutrient molecules in the SI tract is derived and then it is used to formulate the transformation of starch into glucose. Finally, two performance metrics delay and path loss are listed to evaluate model performance.

\subsection{Food Digestion Process as a Communication System}
In the digestive system, consumed food undergoes several intermediate stages before nutrients contained in food are absorbed into the circulatory system as illustrated in Figure \ref{Digest}. The circulatory system  then transports the nutrients and delivers to different parts of the body. The food digestion process is mapped to the components into a MC system as follows:

\begin{figure}[t!]
    \centering
    \vspace{-0.50cm}
    \includegraphics[width = .6\textwidth]{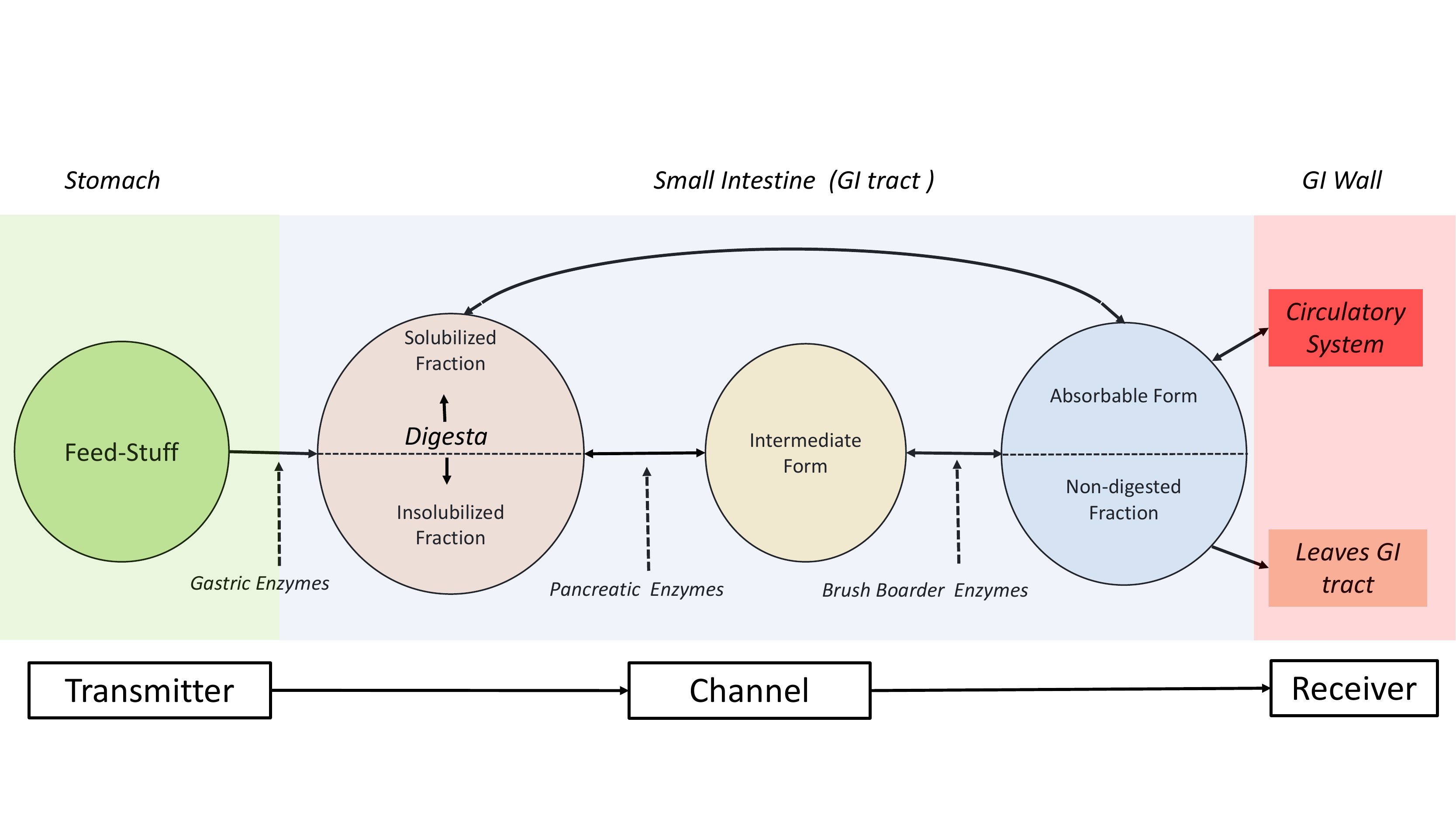}
    \vspace{-0.7cm}
    \caption{An overview of the digestion process as a molecular communication system}
    \label{Digest}
\end{figure}

\begin{itemize}
    \item {\bf Stomach:} is the organ where food is mixed with gastric juice broken down mechanically and chemically before entering into the small intestine. Therefore, the stomach acts as a reservoir of gastric content. 

    \item {\bf Small Intestine (SI tract):} is the organ where most of the chemical digestion takes place due to enzymatic reactions. Its muscles' contraction and relaxation creates convective movement of the nutrient along the SI tract, while their absorption through diffusion. Its wall consists of \emph{villi} that increases the surface area to absorb nutrients effectively.

    \item {\bf Digestion:} is the process by which consumed food converts into absorbable nutrients through enzymatic hydrolysis. Different types of enzymes such as amylases, lipases and proteases help to hydrolyze dietary nutrients such as starch, lipid and protein. In particular, human diet contains a greater portion of carbohydrates of which starch is the main source. Starch is hydrolized into maltose, maltotriose and limit dextrins within the SI tract by pancreatic amylase, while some residual (resistant starch) pass into the large intestine where intestinal bacteria-based digestion takes place. The brush border enzymes (maltase-glucoamylase and sucrase-isomaltase) in the SI tract further break down these intermediate digestion products into glucose.

\end{itemize}
The digestive process described above are characterized and  mapped into an advection-diffusion and reaction based MC system as follows (see Figure \ref{Digest}):

\begin{itemize}
    \item [(a)] {\bf Transmitter:} Stomach acts as a transmitter as it releases digesta into the SI tract.
    \item [(b)] {\bf Channel:} Since the SI tract facilitates digesta to moving along the GI wall providing a medium to convert food particles into absorbable nutrients, the SI transport mechanism is represented as a MC channel.
    \item [(c)] {\bf Receiver:} The SI wall represents the receiver of a MC system as it selectively allows nutrient molecules to penetrate through the SI wall into the circulatory system.  
\end{itemize}


\subsection{Feed Stuff Transport Model}
\begin{figure}[t!]
    \centering
    \includegraphics[width = .45\textwidth]{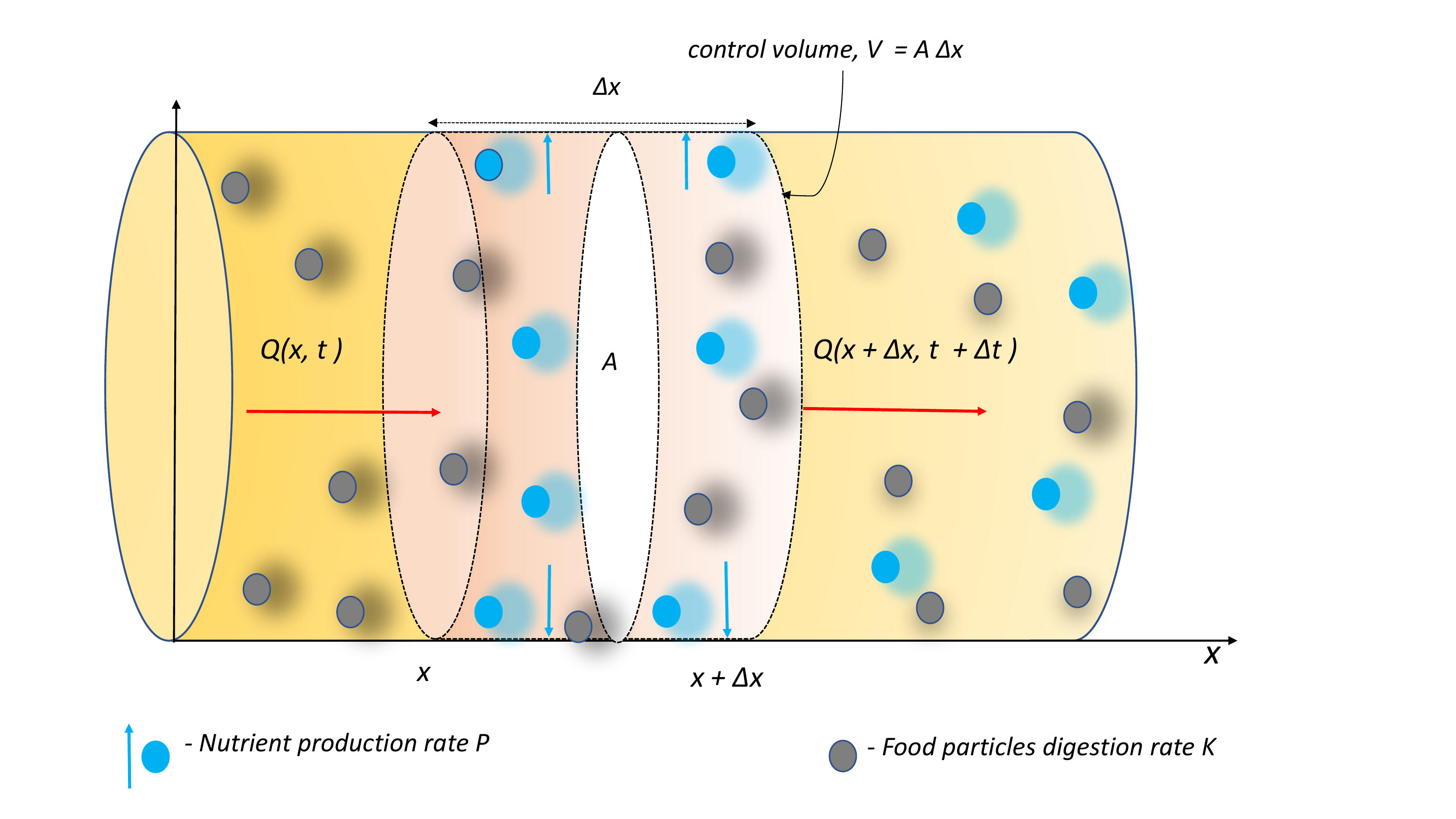}
    \caption{Transport and production of nutrients during the small intestinal digestion of food particles.}
    \label{Transport}
\end{figure}

Figure \ref{Transport} illustrates food particles (in terms of mass) entering the SI tract from the stomach and the rate of mass flow or mass flux $Q (gm^{-2}s^{-1})$ through a cross sectional area $A (m^2)$ in the SI tract. We also consider a control volume of depth $\delta x$ as $V = A\delta x$. The rate of change of the nutrient mass per unit volume (concentration) $C(x,t)$ ($gm^{-3}$) through the control volume $V$ ($m^3$) can be expressed through the following mass balance expression.

\begin{equation}\label{nu-0}
  V\dv{C(x,t)}{t} = \textrm{Mass}_{in} -  \textrm{Mass }_{out},
\end{equation}
where the terms, inlet and outlet mass of nutrients ($\textrm{Mass}_{in}$ and $\textrm{Mass}_{out}$) in equation \ref{nu-0} are formulated as 
\begin{eqnarray}\label{nu-1}
    \textrm{Mass}_{in} &=& Q(x,t)_{in} A + \textrm{Nutrient Production}, \nonumber \\
    &=& Q(x,t)_{in}A + PC(x,t)V, \label{mass-1} \\
    \textrm{Mass}_{out} &=& Q(x+\delta x,t)_{out} A + \textrm{Nutrient Absorption}, \nonumber \\
    &=& Q(x + \delta,t)_{out} A + KC(x,t) V, \label{mass-2}
\end{eqnarray}
where $P$ ($s^{-1}$) and $K$ ($s^{-1}$) are respectively the nutrient production and absorption rates. By substituting equation \ref{mass-1} and \ref{mass-2} in \ref{nu-0}, the change in nutrient mass can be expressed as   
\begin{equation}\label{nu-2}
    \dv{C(x,t)}{t} = -\dv{Q(x,t)}{x} - (K-P)C(x,t).
\end{equation}

Considering the advection-diffusion based flow of nutrients in the SI tract, the mass flow rate of nutrients can be expressed as in equation \ref{nu-3} by using the Fick's first law. 

\begin{equation}\label{nu-3}
    Q(x,t) = C(x,t)u - D \frac{\partial C(x,t)}{\partial x},
\end{equation}
where $u$ ($ms^{-1}$) is the average velocity of the nutrient mass flow. Then, equation \ref{nu-4} expresses the the partial derivative of equation \ref{nu-3} with respect to traveling distance, $x$, and is represented as 

\begin{equation}\label{nu-4}
    \dv{Q(x,t)}{x} = u\frac{\partial C(x,t)}{\partial x} - D \frac{\partial^2 C(x,t)}{\partial x^2}, 
\end{equation}

Combining  equation \ref{nu-2} and \ref{nu-4}, the change in nutrient mass can be expressed as in equation \ref{nu-6}. In general, this is known as the governing equation of the advection-diffusion and reaction based fluid flow in a pipe, and is represented as 

\begin{equation}\label{nu-6}
    \frac{\partial C(x,t)}{\partial t} + u(x,t) \frac{\partial C(x,t)}{\partial x} = D \frac{\partial^2 C(x,t)}{\partial x^2} - (P-K)C(x,t).  
\end{equation}
Models from \cite {17} are used to compute the $P$ and $K$. The nutrient production rate is derived using the Michaelis-Menten-kinetics and this is based on $P =V_{max} \frac{[S]}{K_{max} + [S]}$, where $V_m$ is the maximum reaction rate and $K_{max}$ is the half saturation concentration. The nutrient absorption rate is derived by using the {\emph Sherwood number}, {\emph Reynolds number} and {\emph Schmidt number} as   $K = \frac{2}{r_m} f K_a$, $K_a = 1.62 \left(\frac{u D^2}{Ld}\right)^{\frac{1}{3}} $, $D = \frac{K_BT}{6\pi \mu r_m}$, where, $D$ is the diffusion coefficient, $K_B$ is the Boltzmann constant, $T$ represents the absolute temperature and $r_m$ is the radius of nutrient molecules and $L$ and $d$ are the length and diameter of the SI tract, respectively.
\subsection{Starch Digestion Model}
Following equation \ref{nu-6}, starch digestion into glucose and absorption into blood in the SI tract are formulated below as a system of differential equations \ref{eq-1}-\ref{eq-3}. In this model, it is assumed here that the starch digestion starts after entering the SI tract. Also, the model is created using carbohydrate as the only source of nutrients, but humans and animals consume mixed macro-nutrients within diverse food matrices, which impact on the how the carbohydrates are digested. Thus, the presence of other macro-nutrients in the carbohydrate enriched diet significantly impacts on the glucose availability in the blood.

In response to the starch in the stomach $C_{st}(t)$ (sometimes the calls gastric volume) emptying at a rate $\gamma C_{st}$ (see equation \ref{eq-1}, where $\gamma = \log(2)/t_{1/2}$ and $t_{1/2}$ is the half-gastric emptying time), will result in the starch concentration in the SI tract $C_s(x,t)$ increasing at the same rate. While starch molecules travel along the SI tract under the advection-diffusion mechanism, they react with the enzymes  and consequently produces starch hydrolysis products (e.g., maltose) and then brush border enzymes (e.g., maltases) convert them into glucose. This study does not take into account concentrations of intermediate products and considers only the conversion of starch into glucose.  Hence, the $C_s(x,t)$ decreases along the SI tract at a rate  $V_m\frac{C_s}{K_m+C_s}$ (see equation \ref{eq-2}). The glucose in the SI tract, $C_g(x,t)$, then increases at the same rate as the starch digestion. At the same time, it decreases at a rate $KC_g(x,t)$ due to the absorption of glucose molecules into the circulatory system while they travel along the SI tract under the same mechanism as the starch molecules (see equation \ref{eq-3}). Please see \cite{18} for more information about the digestive system modeling. Altogether, the system of differential equations to characterize the starch digestion and production of glucose is expressed as

\begin{eqnarray}\label{di-eqn-new}
\dv{C_{st}(t)}{t} &=& -\gamma C_{st}(t), \label{eq-1}\\
\frac{\partial C_s(x,t)}{\partial t} &=& D \frac{\partial^2 C_s(x,t)}{\partial x^2} - u \frac{\partial C_s(x,t)}{\partial x} - V_m\frac{C_s(x,t)}{K_m+C_s(x,t)} + \gamma C_{st}(t), \label{eq-2}\\
\frac{\partial C_g(x,t)}{\partial t} &=& D \frac{\partial^2 C_g(x,t)}{\partial x^2} - u \frac{\partial C_g(x,t)}{\partial x} + V_m\frac{C_s(x,t)}{K_m+C_s(x,t)} - K C_{g}(x,t). \label{eq-3}
\end{eqnarray}
\subsection{Analytical Solution}
Based on the differential equations \ref{eq-1}-\ref{eq-3}, we derive analytical solutions  assuming $K_m << C_s(x,t)$. In reality, this assumption emphasizes that nearly all enzymes will be occupied by the starch molecules. Hence, this assumption leads to a simplification that $V_m\frac{C_s}{K_m + C_s} \approx V_m$. The analytical solution of three equations \ref{eq-1}-\ref{eq-3} are given as follows
\begin{itemize}
    \item [(1)] {\bf Starch in Stomach:} Suppose at $t=0$, $C_0$ amount of starch is consumed, then the solution of equation \ref{eq-1} represents the variability in starch concentration in the stomach, $C_{st}$ over time, and is represented as  
    \begin{equation} \label{eq-1-sol}
        C_{st}(t) = C_0 e^{-\gamma t}.
    \end{equation}
    \item [(2)] {\bf Starch in SI tract:} The solution for equation \ref{eq-2} is represented as
        \begin{equation} \label{eq-2-b}
            C_s(x,t) = \frac{f_s}{\sqrt{4\pi Dt}}e^{\frac{-(x-ut)^2}{4Dt}} -C_0e^{- \gamma t} - V_mt.
        \end{equation}
        where $f(x,t) = \left[C_s(x,t)+ C_0e^{-\gamma t} +V_mt\right] e^{ \left( \frac{u^2}{4D}t -\frac{u}{2D}x \right) } $ and $f_s = f(0,0)$. 
        
    \item [(3)] {\bf Glucose in SI tract:} Following the similar approach as in the step 2, the solution of equation \ref{eq-3} can be expressed as follows.  
        \begin{equation} \label{eq-3-b}
            C_g(x,t) = \frac{f_g}{\sqrt{4\pi Dt}}e^{\frac{-(x-ut)^2}{4Dt} -Kt} + V_mt,
        \end{equation}
        where $f(x,t) = \left[C_g(x,t)-V_mt\right] e^{ \left[ \left( \frac{u^2}{4D} + K\right)t -\frac{u}{2D}x \right] } $ and $f_g = f(0,0)$.
\end{itemize}
\subsection{Channel Characteristics}
Two performance metrics listed below are used to evaluate the molecular communication system of the starch digestion dynamics in the SI tract.
\begin{itemize}
    \item [(1)]{\bf Path Loss (PL):} This is considered as the loss of starch or glucose traveling from an arbitrary distance $x_1$ to $x_2$ ($0 \leq x_1 < x_2 \leq l$) along the SI tract. By following the path loss equation given in \cite{10}, PL is computed as 
        \begin{equation}\label{pl}
            PL = 10 \log_{10}\left( \frac{C_{g}(x_1,t)}{C_g(x_2,t)}\right).
        \end{equation}
    \item [(2)] {\bf Delay:} It is generally assumed that a significant amount ( $>90\%$) of consumed carbohydrates enters into the circulatory system as glucose within $2-3$ hours. In this study, we assumed that $1.5$ hours as the benchmark to digest $>50\%$ of consumed carbohydrates, the delay is computed as the difference between the time taken to digest and then absorb consumed carbohydrates and the benchmark time.
\end{itemize}
\section{Results}\label{res}
The model derived in the previous section is used here to characterize starch digestion dynamics along the SI tract through simulations studies which were performed by using the model parameters listed in Table \ref{tab-1}. 

\begin{table}[t!]
\centering
\caption{Model Parameters Used for Simulations.}
\begin{tabular}{|l|c|l|}
\hline
Parameter & Symbol & Value  \\
\hline
\hspace{.25cm} Small intestine length & $l$ & $ 6.9 m$ (male) \\
\hspace{.25cm} Small intestine diameter & $d$ & $2.5 cm$  \\
\hspace{.25cm} Viscosity & $\mu$  & $0.01 - 10 Pa s$ \\
\hspace{.25cm} Radius of Glucose molecules & $r_m$ & $ 0.38 nm$\\
\hspace{.25cm} Surface area increase due to fold, vili and microvili & $f$ & $ 12$\\
\hspace{.25cm} Mean velocity & $u$          & $ 1.7 \times 10^{-4} ms^{-1} $ \\
\hspace{.25cm} Maximum reaction rate & $V_{max}$          & $ 25 mM min^{-1} $ \\
\hspace{.25cm} Half saturation concentration & $K_{max}$          & $ 9 mM $ \\
\hspace{.25cm} Half gastric emptying time & $t_{1/2}$          & $ 1 hour$\\

\hline
\end{tabular}\label{tab-1} 
\end{table} 
\subsection{Starch Digestion and Production of Glucose }

\begin{figure}[t!]
    \centering
    \includegraphics[width = .7\textwidth]{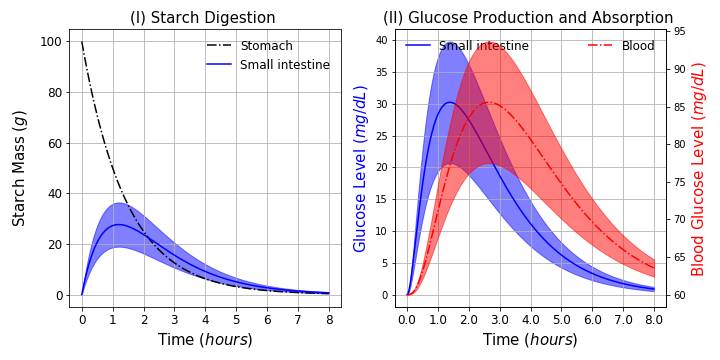}
    \caption{Having consumed 100$g$ of carbohydrates, change in starch and glucose concentrations in the stomach as SI tract.}
    \label{fig-1}
\end{figure}

As mentioned in the previous section, starch is converted into glucose while traveling along the SI tract so that both the starch and glucose concentrations vary over the SI tract with respect to the distance that they travel  over time. Figure \ref{fig-1} shows the variability in the starch concentrations averaged over distance in both the stomach and SI tract along with the glucose concentration, given that $100 g$ of carbohydrates is consumed and the normal blood glucose level is $60 mg/dL$. As Figure \ref{fig-1}$(I)$ depicts, starch mass in the stomach decreases over time and a greater fraction of consumed carbohydrates has been entered into the SI tract within $1-1.5 hours$. As a consequence, initially, the starch concentration in the SI tract depicts an increase and then it gradually decreases in response to the degradation of starch into glucose. As a result, in Figure \ref{fig-1}$(II)$, the glucose concentration in the SI tract increases and the maximum increase is around $30 mg/dL$. In response to the absorption of glucose into blood, the blood glucose level increase and then decreases as glucose is stored in the body. The Beragman's minimal model \cite{21} was used to compute the blood glucose level.  

In general, Figure \ref{fig-1} shows a greater portion ($ > 90\%$) of starch has been digested and the produced glucose has been absorbed into the blood around in $2-3$ hours and the observed maximum glucose change is around $30 mg/dL$.  
\subsection{Effectiveness of Starch Digestion}

\begin{figure}[t!]
\centering
\begin{subfigure}{.75\textwidth}
  \centering
  \includegraphics[width= 1\linewidth]{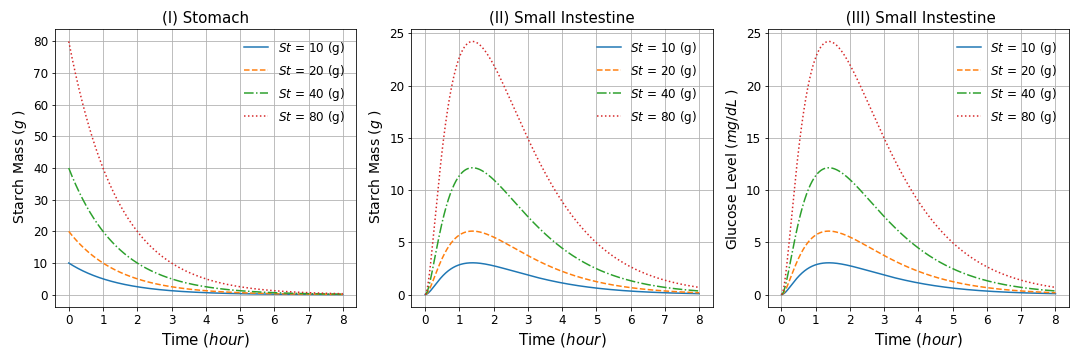}  
  \caption{}
  \label{fig-2-1}
\end{subfigure}
\begin{subfigure}{.75\textwidth}
  \centering
  \includegraphics[width=1\linewidth]{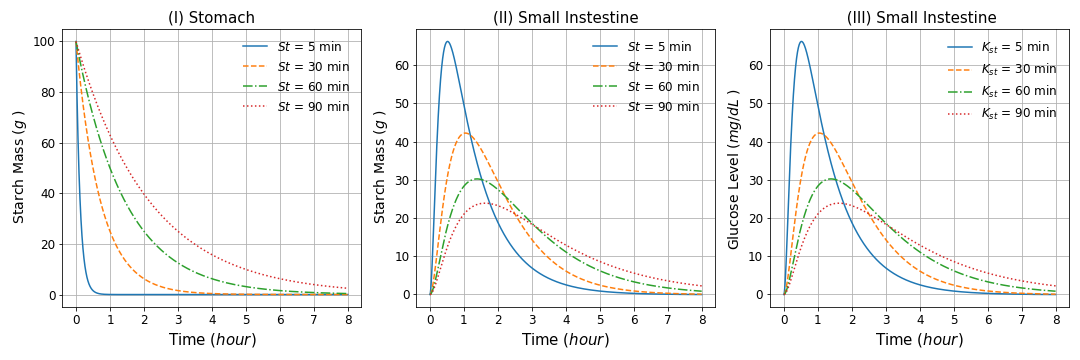}  
  \caption{}
  \label{fig-2-2}
\end{subfigure}
\begin{subfigure}{.75\textwidth}
  \centering
  \includegraphics[width= .7\linewidth]{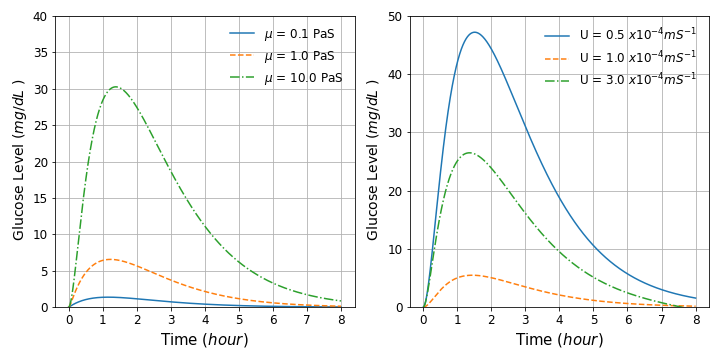}  
  \caption{}
  \label{fig-2-3}
\end{subfigure}

\caption{ Variability in starch and glucose concentration in the stomach and SI tract with respect to the (a)  amount of consumed carbohydrates, (b) and the half stomach emptying time, and (c) change in viscosity ($\mu$) and velocity ($u$)(only in the SI tract).}
\label{fig-2}
\end{figure}

The effectiveness of the starch digestion process could, however, be affected by a number of factors. This results in delay in production and absorption of glucose which is used as the main source of energy required for different body functions. Figure \ref{fig-2} shows the impact of selected parameters (consumed quantity, stomach emptying time, viscosity and velocity) on the starch digestion process. According to Figure \ref{fig-2-1}, the time taken to absorb the produced glucose into blood is slightly increasing with the increase in consumed quantity of carbohydrates. On the other hand, Figure \ref{fig-2-2} depicts that the time taken to digest $100g$ of carbohydrates and then absorbing the produced glucose into blood is higher with longer stomach emptying time. In addition, Figure \ref{fig-2-3} shows the change in glucose concentration in the SI tract in response to the increase in velocity ($u$) and viscosity ($\mu$) of glucose molecules in the SI tract, given that $100g$ of carbohydrates has been consumed. In the model, the increase in viscosity results in increasing glucose level as  higher viscosity slows down the starch movement speed and this in turn, allows starch molecules to spend long period of time in the SI tract. Conversely, increasing velocity decreases the time spent by starch molecules in the SI and hence, results in decrease of  the glucose level. 
\subsection{Delay in Starch Digestion} 

\begin{figure}[t!]
    \centering
    \includegraphics[width = .85\textwidth]{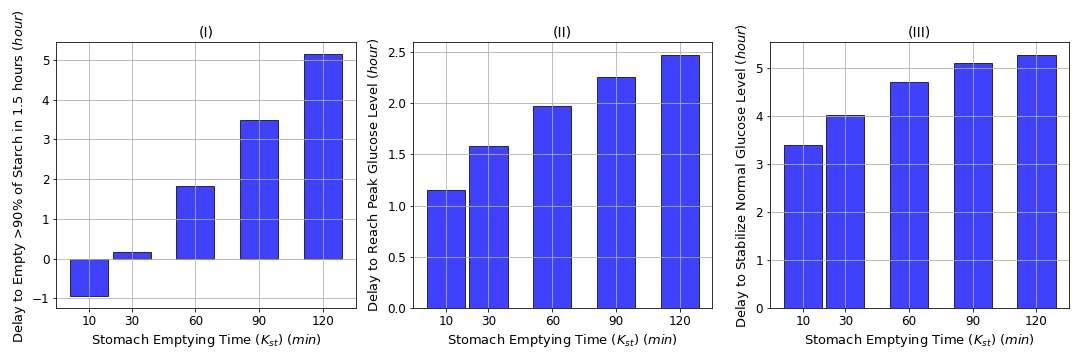}
    \caption{The impact of stomach emptying time on the starch digestion delay and corresponding delay it creates in stabilizing blood glucose level. }
    \label{fig-3}
\end{figure}

For a healthy person, it is generally known that carbohydrates digestion results in a sudden increase in blood glucose level and on average takes an hour to reach its peak and then it stabilizes in response to the secretion of insulin from the pancreas.  Figure \ref{fig-2} showed that  different factors can have a significant influence on delaying the digestion process. The stomach emptying time acts as one of the most influential factors among them because long stomach emptying time can result in significant delay in the remaining digestion process. So, this study explored only the impact of the stomach emptying time and Figure \ref{fig-3} displays the impact of this on delaying the starch digestion process. Given the average time taken to digest  $>50\%$ of consumed carbohydrates is $1.5 hours$, Figure \ref{fig-3}(I) shows the delay of starch entering from the stomach into the SI tract with increasing stomach emptying time. The corresponding increase in time to reach maximum blood glucose level and reaching back to normal level (stabilizing glucose level) is displayed respectively in Figure \ref{fig-3}($II$) and \ref{fig-3}($III$). Therefore, as seen in Figure \ref{fig-3}, longer stomach emptying time delays the starch digestion process, which in turn, could cause delaying the body glucose demand required to produce energy. Consequently, such a delay may lead to certain disorders in different body functioning. 
\subsection{Path loss with respect to Starch and Glucose}

\begin{figure}[t!]
\begin{subfigure}{.5\textwidth}
  \centering
  \includegraphics[width= 1\linewidth]{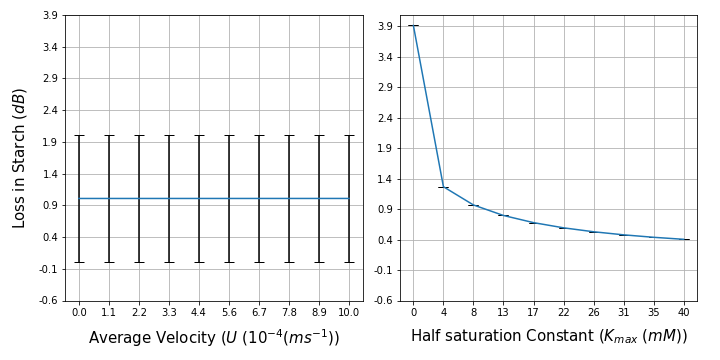}  
  \caption{}
  \label{fig-4-1}
\end{subfigure}
\begin{subfigure}{.5\textwidth}
  \centering
  \includegraphics[width=1\linewidth]{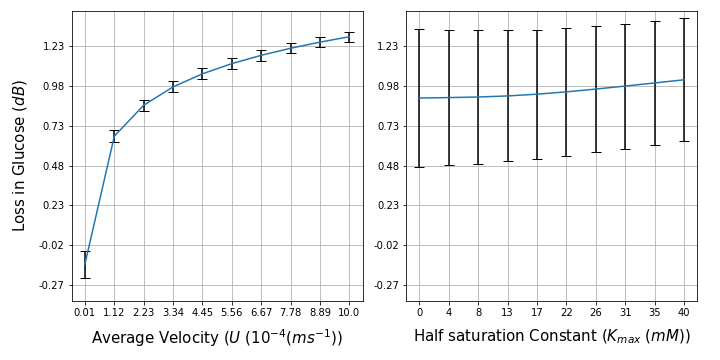}  
  \caption{}
  \label{fig-4-2}
\end{subfigure}

\caption{Path loss with respect to starch and glucose while varying velocity $u$ and $K_{max}$ together.}
\label{fig-4}
\end{figure}

\begin{figure}[t!]
\begin{subfigure}{.5\textwidth}
  \centering
  \includegraphics[width= 1\linewidth]{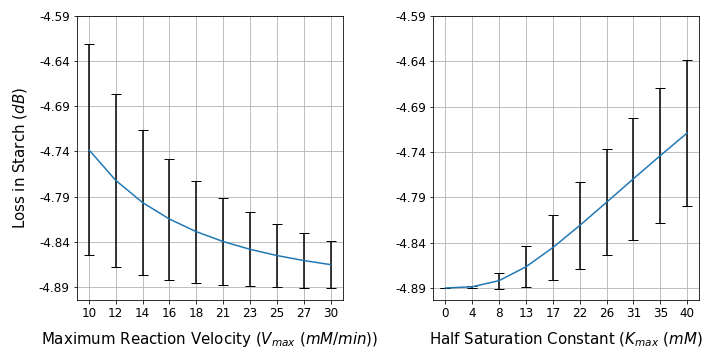}  
  \caption{}
  \label{fig-5-1}
\end{subfigure}
\begin{subfigure}{.5\textwidth}
  \centering
  \includegraphics[width=1\linewidth]{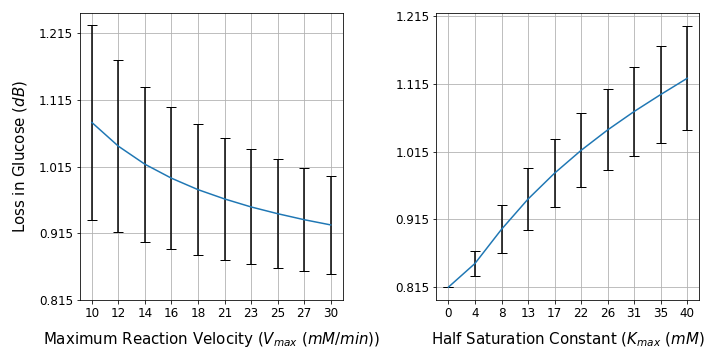}  
  \caption{}
  \label{fig-5-2}
\end{subfigure}

\caption{ Path loss with respect to starch and glucose while varying maximum reaction rate $V_{max}$ and half saturation concentration $K_{max}$ together.}
\label{fig-5}
\end{figure}

The PL can result from both the loss in undigested starch  and unabsorbed glucose in the small intestine. Figure \ref{fig-4} shows the variability in PL in response to the change in velocity $u$ and half saturation concentration $K_{max}$ over the range of $[0.01-10]\times 10^{-4} m/s$ and $[0-40]mM$, respectively. As displayed in Figure \ref{fig-4-1}, the PL with respect to starch does not show significant change with increasing velocity, but increasing $K_{max}$ contributes to decreasing PL at a declining rate. PL with respect to glucose given in Figure \ref{fig-4-2} shows an increase with increasing both $u$ and $K_{max}$ and also the increasing rate with $u$ is greater compare to that with $K_{max}$, but the increment is declining with larger $u$.  Besides, Figure \ref{fig-5} shows the impact of $V_{max}$ and $K_{max}$ together on PL is depicted while $V_{max}$ and $K_{max}$ are varying over the ranges $[10-30] mM/min$ and $[0-40] mM $, respectively. The PL with respect to both starch and glucose shows decreasing trend with respect to increase in $V_{max}$, but they show an increasing trend with increasing $K_{max}$. 

\section{Discussion}\label{dis}
Exploring starch digestion dynamics is important for a number of reasons. It is the main source of carbohydrates in our diet and acts as the highly efficient energy source required for body function. Thus, efficient carbohydrate digestion is vital, for instance in intensive exercises as it avoids the body using protein stored in the internal organs to produce energy. This in turn helps preserving lean muscle mass and the use of dietary protein for other important purposes such as building muscles rather than consuming for generating energy. Moreover, carbohydrate helps the function of brain and nervous system  and metabolism of fat. Effectiveness of starch digestion could, however, be varied from person to person due to variability in influenced factors such as gastric emptying times, flow of digestive content, digestive capacity and efficacy of glucose transporters, as well as the glucose absorption into the blood \cite{12}. Thus, any abnormality of one or more of these factors could potentially lead to creating potential disorders in aforementioned various body functions interconnected with glucose. For instance, imbalance in optimum blood glucose level could cause a number of issues because high plasma glucose level could leads to microvascular  and neural damages, and renal inefficiency while low glucose level causes anxiety, aggressiveness, coma and eventually death \cite{13}. Therefore, efforts to understand influence of these factors would be therapeutically important. 

The MC-based approach proposed can provide insights which can contribute to characterizing the influence of such factors on the starch digestion process effectively. It would be useful for taking actions to control diabetes and obesity, for example. Therefore, this section discusses the significance of these insights for the transmitter, channel and receiver through  simulation studies that can contribute to design of appropriate interventions. 
\subsection{Transmitter: Effective Gastric Emptying Time}
In the digestion system, effectiveness of functions associated with the transmitter (i.e., stomach) is a crucial entity that can control the whole starch digestion process. This is because the glucose absorption into blood depends on the rate at which gastric content is released into the small intestine. Since the gastric emptying rate $\gamma$ is computed by using the time to empty the stomach content by half of its volume, it directly influences on the delay in the whole starch digestion process. Various factors influence the increasing  gastric emptying rate and this can include high volume, viscosity, exercise level and relative nutrient contents in the meal. Moreover, aging also increases this value, which in turn, decreases the efficacy of absorption of glucose and also the calcium and iron too \cite{14}. These dynamics can easily be characterized by using the MC-based approach as it allows exploring variability in starch and glucose concentrations in the small intestine with respect to change in certain model parameters. For instance, both Figure \ref{fig-2} and \ref{fig-3} shows the potential of the proposed model in exhibiting the impact of variability in delay with increasing stomach emptying time, velocity and viscosity. Increasing delay in starch digestion could directly influence the homeostasis in blood glucose level which is crucial for optimum body functions. Therefore, these insights could be incorporated for exploring dynamics in insulin sensitivity and glucose effectiveness which are very important for controlling blood glucose level, as well as designing personalized treatments for diabetes patients.
\subsection{Channel: Efficacy in Mechanical and Chemical Functions}
Digestive dynamics within the small intestine can primarily be characterized by its mechanical and chemical functionalities. The mechanical functionalities include movements in muscles while enzymatic reactions include the chemical functionalities. Variability in these processes can thus impact effectiveness of the digestion process. 

The contraction and relaxation of muscles in the SI tract help effective separation, mixing and propelling of digesta, which in turn, creates convective flows allowing effective transformation of digesta into absorbable nutrients. The Brownian motion of nutrient molecules creates diffusion-based movement which helps nutrient absorption into the blood. Optimal functionality in these processes allows sufficient transient time that is essential for complete digestion and  increasing the nutrients absorbed into blood. In the model discussed here, bulk flow velocity ($u$) and viscosity ($\mu$) are two key parameters which can make a greater influence on the movement of nutrient molecules. For this reason, Figure \ref{fig-2-3} shows a greater glucose production with low velocity and high viscosity. Figure \ref{fig-5-1} also supports this observation, indicating a lower path loss with respect to smaller velocity.   

With regards to the chemical reactions, the model considered only the enzymatic action on the starch molecules based on Michaelis-Menten kinetics. Due to the complexity in solving the differential equations, the analytical solutions were derived considering the half saturation concentration ($K_{max}$) is very small compared to the starch concentration. This means  all enzymes will have been occupied by starch molecules by the saturation stage of the reaction between the starch and enzymes. Therefore,  Figure \ref{fig-4} shows opposite trends in loss with respect to the starch and glucose with increasing $K_{max}$. Moreover, Figure \ref{fig-5} suggest that smaller half saturation concentration results in greater contribution to reduced path loss. 
\subsection{Receiver: Effective Nutrient Absorption}
Lower path loss with respect to both starch and glucose can contribute to increasing the amount of nutrients detected by the receiver (i.e., the small intestine wall). According to the model, nutrient absorption also depends on bulk flow size of nutrient molecules, velocity, and viscosity as parameters related to small intestine such as length and diameter are fixed. Therefore, as discussed above, short gastric emptying time, low velocity and high viscosity can contribute to improving the glucose absorption effectiveness. The impact of nutrient molecule size was not considered since this study considered glucose absorption only.
\subsection{Future Directions}
Since the study lays a strong foundation to explore the digestive dynamics in terms of MC concepts, it unveils a number of ways that this study could be further extended. Most importantly, given the amount of glucose absorbed into the blood, variability in blood glucose  level in the body can be explored with respect to a range of factors such as age, gender, and physical fitness. This can help in terms of designing therapeutic treatments for chronic diseases such as diabetes. Moreover, it would also help in taking actions against obesity as controlling carbohydrates intake that can effectively stimulate burning excess body fat by increasing activity levels. In addition, males and females differ in their food and calorie intake, with males generally consuming more food including carbohydrates than females. The model discussed here is general and not specific to any gender, but there is a potential of using it to explore gender-based variability in digestive dynamics by including the gender related factors into the model parameters, such as the gut length and need of calories.

This study accounts the impact of few parameters on the starch digestion process so that it is able to provide an overall idea about the digestion dynamics. This is because the enzymatic activity depends on the $pH$ value and the temperature but the study assumes they do not vary over the digestive tract though they have a significant impact on the digestion process. The model assumes the gut width is fixed but it varies along the tract, and by the development stage as the individual grows, so does the size of the gut including the width and length, which can accommodate increased intake to support growth. 

Also the starch digestion starts in the mouth. However, this study assumed the starch stays intact until it reaches the small intestine. The model was created using carbohydrate as the only source of nutrients, but humans and animals consume mixed macronutrients within diverse food matrices, which impact on the how the carbohydrates are digested. The presence of other macronutrients in the carbohydrate enriched diet significantly thus impacts on the glucose availability in the blood. Taking into account these factors in the model development process will also contribute to improving the potential of creating an accurate digestive dynamics digital model. 

\section{Conclusion}\label{con}
This study looks at the digestive system functionality as a MC system and then proposes an advection-diffusion and reaction based model to characterize digestion system dynamics in the context of starch digestion. Based on communication system evaluation, two performance metrics, which are delay and path loss are used to explore the influence of different properties of the digestive system in the context of starch digestion. According to the observations derived from simulation of an analytical model, the gastric emptying rate, velocity and viscosity of digesta along with the parameters related to enzymatic activity on the starch such as half saturation concentration  influences on the delay and path loss in the starch digestion process. These observations primarily suggest that shorter gastric emptying time with low velocity and high viscosity of digesta can contribute to improving the amount of glucose absorbed into the blood. Therefore, these insights help in expanding the knowledge about digestive system dynamics so that they can contribute towards the development of novel solutions for future food-related global crisis such as obesity and food waste. Most importantly, the study lays a concrete foundation to drive digestive system studies towards a new direction that will enable us to create a digital twin that will personalize our ingredients that suits our personal physiological settings as well as internal digestion functions. 

%
%
\section*{Acknowledgment}
This research was supported by Science Foundation Ireland and the Department of Agriculture, Food and Marine on behalf of the Government of Ireland VistaMilk research centre under the grant 16/RC/3835. 

\bibliographystyle{splncs04}
\bibliography{Paper}


\end{document}